\documentclass[aps,prl,twocolumn,showpacs,bibnotes]{revtex4}

\usepackage{graphicx}
\usepackage{amssymb}
\usepackage{bm}

\newcommand{\be}{\begin{equation}}
\newcommand{\ee}{\end{equation}}

\newcommand\pictc[5]{\begin{figure}
                   \centerline{
                   \includegraphics[width=#1\columnwidth]{#3}}
               \protect\caption{\protect\label{fig:#4} #5}
                \end{figure}            }

\newcommand\pict[4][1]{\pictc{#1}{!tb}{#2}{#3}{#4}}
\newcommand\rpict[1]{\ref{fig:#1}}

\newcommand\leqt[1]{\protect\label{eq:#1}}
\newcommand\reqtn[1]{\ref{eq:#1}}
\newcommand\reqt[1]{(\reqtn{#1})}

\newcounter{Fig}

\begin{document}
\begin{sloppy}
\title{Soliton dynamics in deformable nonlinear lattices}

\author{Andrey A. Sukhorukov}
\affiliation{Nonlinear Physics Centre and Centre for Ultra-high bandwidth Devices for Optical Systems (CUDOS),
Research School of Physical Sciences and Engineering,
Australian National University,
Canberra, ACT 0200, Australia}

\begin{abstract}
We describe wave propagation and soliton localization in photonic lattices which are induced in a nonlinear medium by an optical interference pattern, taking into account the inherent lattice deformations at the soliton location. We obtain exact analytical solutions and identify the key factors defining soliton mobility, including the effects of gap merging and lattice imbalance, underlying the differences with discrete and gap solitons in conventional photonic structures.
\end{abstract}

\pacs{42.65.Tg,  
      42.65.Jx,  
}

\maketitle

The effect of Bragg scattering from a periodic potential is a fundamental phenomenon which  is responsible for a strong modification of wave dispersion and the appearance of spectral gaps. The structure of band-gap spectrum in crystals defines the electron transport properties, and similar concepts were successfully developed in the field of optics~\cite{Joannopoulos:1995:PhotonicCrystals}. The ultimate flexibility in managing wave transport and localization may be achieved in dynamically induced lattices. Such reconfigurable lattices can be realized in any nonlinear media, where a modulated wave can modify the medium characteristics (e.g. an optical refractive index~\cite{VanSimaeys:2004-223902:PRL, Neshev:2004-486:OL, Martin:2004-123902:PRL}) and induce an effective periodic potential. 

Nonlinearity also supports wave localization inside the band-gaps in the form of discrete and gap solitons, when dispersion or diffraction is suppressed through self-focusing~\cite{Eggleton:1999-587:JOSB, Christodoulides:2003-817:NAT, Mandelik:2004-93904:PRL, Fleischer:2003-147:NAT, Neshev:2004-83905:PRL}. The dynamics of such solitons in fixed lattices is strongly influenced by a self-induced Peierls-Nabarro (PN) potential which can result in soliton trapping, and this effect has applications for intensity-dependent beam steering~\cite{Morandotti:1999-2726:PRL}.
Recent theoretical studies have indicated that the PN potential can vanish in nonlinear lattices which become deformed at the soliton location~\cite{Desyatnikov:2003-153902:PRL, Desyatnikov:2004:ProcNLGW+}, however soliton trapping was observed in recent experiments due to soliton-lattice interaction~\cite{Neshev:2004-486:OL, Martin:2004-123902:PRL}. In this Letter, we describe the key mechanisms which determine the soliton mobility in deformable nonlinear lattices when the PN potential is absent. In particular, we identify the fundamental effect of band-gap merging, and suggest how it can be controlled through the lattice imbalance.

We will analyze the case when the interaction between the localized beam and the nonlinear lattice is phase-insensitive. This situation is realized for optical waves which are mutually incoherent~\cite{Kang:1996-3699:PRL, Neshev:2004-486:OL, Martin:2004-123902:PRL}.
We consider the (1+1)D geometry, where waves can diffract in one spatial dimension, and they are confined by a guiding potential in the other transverse dimension.
Then, the wave dynamics can be approximately described by the coupled nonlinear Schr\"odinger (NLS) equations for the normalized wave envelopes $E_n$,
\begin{equation} \leqt{NLS}
   i \frac{\partial E_n}{\partial z}
   + \frac{ \partial^2 E_n}{\partial x^2}
   + 2 \sigma I E_n
   = 0,
\end{equation}
where $x$ is the transverse spatial coordinate, $z$ is the propagation direction, $I=|E_1|^2 + |E_2|^2$ is the total intensity, and $\sigma=\pm 1$ stands for the focusing or defocusing nonlinearity, respectively.  We have neglected the higher-order nonlinear effects in Eq.~\reqt{NLS} to identify the most general physical effects.
The model Eqs.~\reqt{NLS} are fully integrable~\cite{Manakov:1973-505:ZETF}, and their solutions can be obtained analytically. 

We first summarize the properties of nonlinear lattices which are found as stationary periodic solutions~\cite{Carr:2000-63610:PRA+} of Eq.~\reqt{NLS} in the most general form, $E_1 = r_1(x) e^{i \varphi_1(x) + i \beta_1 z}$. Here the amplitude [$r(x)$] and phase [$\varphi(x)$] profiles are fixed, and the propagation constant $\beta_1$ is proportional to the wavevector component along the $z$ direction. We can present the lattice intensity profile in the same form for both the cases of self-focusing ($\sigma=+1)$ and self-defocusing ($\sigma=-1)$ nonlinearities,
\begin{equation} \leqt{lattice}
   \sigma r_1^2(x) = A^2 {\rm cn}^2( x \kappa, m ) + V_0, 
\end{equation}
where ${\rm cn}$ is the Jacobi elliptic function with modulus $m$ ($0<m<1$), $V_0 = \beta_1 / 3 - \kappa^2 (2 m - 1) / 3$, $A = \kappa \sqrt{m}$, $\kappa = 4 K(\sqrt{m}) / (2 d)$, $K$ is the complete elliptic integral of the first kind, and $d$ is the lattice period. The lattice phase is given in an integral form, 
   $\varphi_1(x) = \int C r_1^{-2}(x)\;dx$, 
where $C^2 = - V_0 ( V_0 + A^2 ) ( V_0 + A^2 - \kappa^2 )$.
The lattice is defined by the following parameters: period $d$, modulus $m$, and the propagation constant $\beta_1$, which should be chosen to satisfy the conditions $A^2 > 0$ and $C^2 \ge 0$. 
The nonlinear waves exhibit strong modulational instability when $\sigma =+1$ and $V_0 \simeq \kappa^2 (1-m)$. In the following we consider the lattices with parameters $\sigma=+1$, $V_0 \simeq 0$ and $\sigma=-1$, $V_0 \simeq -m \kappa^2$, when only weak oscillatory instabilities may appear, which are suppressed in media with saturable nonlinearity~\cite{Aleshkevich:2001-257:QE, Desyatnikov:2003-153902:PRL}.
Such solutions can be excited experimentally by two interfering waves representing the dominant Fourier harmonics,
$E_1(x,z=0) \simeq F_+ \exp( i \pi x / d ) + F_- \exp( -i \pi x / d )$.
The balanced excitation ($F_-=F_+$) produces a lattice with $C=0$ and a flat-phase profile [see Figs.~\rpict{lattice}(a,b)], and imbalance leads to nontrivial phase modulation with $C \ne 0$ [Fig.~\rpict{lattice}(c)].

\pict{fig01}{lattice}{
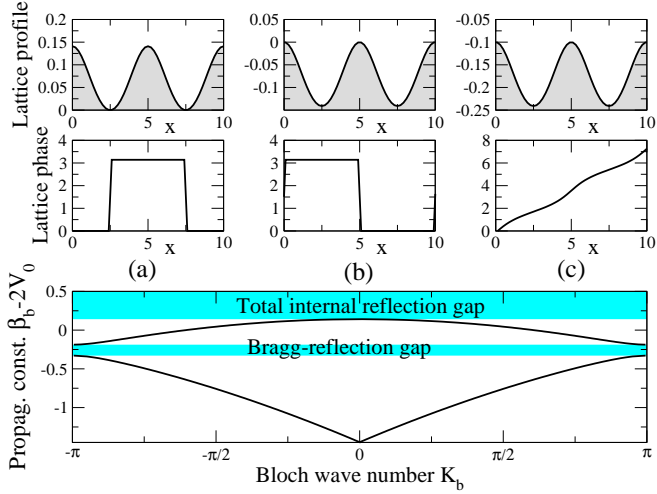
Top rows: Characteristic profiles of nonlinear lattices with $d=5$, $m=0.3$, and the same profiles $\sigma r_1^2(x)$ (top), but different nonlinearities and phase structures: 
(a)~flat-phase in a self-focusing medium ($C=0$,  $V_0=0$, $\sigma=+1$),
(b)~flat-phase in a self-defocusing medium ($C=0$,  $V_0=-A^2$, $\sigma=-1$),
and (c)~nontrivial phase modulation in a self-defocusing medium ($C \simeq 0.12$, $V_0=-A^2-0.1$, $\sigma=-1$). 
Bottom: Linear Bloch-wave dispersion, the same for all the lattices (a-c).
}

The periodic modulation induced by a nonlinear wave strongly affects dynamics of the probe beam which can exhibit Bragg scattering from the lattice. Since the beam-lattice coupling is phase-insensitive, the propagation of small-amplitude probe beam (with an envelope $E_2$) is governed by linear equation with a stationary periodic potential defined by the unperturbed lattice intensity profile [Eq.~\reqt{lattice}], and therefore does not depend on the lattice imbalance. Linear wave dynamics in a periodic potential can be analyzed by decomposing the wave packet into a superposition of extended eigenmodes called Bloch waves which 
are found as solutions of the wave equation in the form $E_2(x,z) = B(x) \exp(i \beta_b z + K_b x / d)$, where $K$ is the normalized Bloch wave number and $B(x) = B(x+d)$ is a periodic wave profile. We immediately notice that nonlinear periodic waves [$E_1(x,z)$]
satisfy the Bloch condition. Moreover, for fixed parameters $d$ and $m$, the lattice intensity profiles are exactly the same [up to a constant shift defined by the value of $V_0(\beta_1)$, see Fig.~\rpict{lattice}(a-c)], and these solutions form a full set of Bloch waves which wave-numbers are found as $K = \varphi(d) - \varphi(0)$. Conversely, for all lattices with particular $d$ and $m$ but different phase structure and type of nonlinearity, the dispersion of Bloch waves $\beta_b(K)$ is equivalent, see Fig.~\rpict{lattice}(bottom). There exists a Bragg-reflection gap for a range of propagation constants $\kappa^2 (m - 1) + 2 V_0 < \beta_b < \kappa^2 (2 m - 1) + 2 V_0$, and a semi-infinite gap due to total internal reflection for $\beta_b > \kappa^2 m + 2 V_0$. 

In order to uncover the fundamental features of nonlinear wave transport in deformable lattices, we analyze the properties of stationary and moving solitons, consisting of nonlinearly coupled spatially localized and extended lattice components. We consider the most general case of phase-modulated lattices with internal imbalance ($C \ne 0$), and identify new effects in comparison with solitons in flat-phase lattices~\cite{Desyatnikov:2003-153902:PRL, Desyatnikov:2004:ProcNLGW+}. We use the approach based on Darboux transformation~\cite{Desyatnikov:2004:ProcNLGW+} to ``add'' a localized soliton to a periodic lattice. In this method, first one calculates auxiliary functions as solutions of equations associated with the Lax pair representation of the integrable Eqs.~\reqt{NLS},
\begin{equation} \leqt{psi}
   \begin{array}{l} {\displaystyle
      \frac{\partial \psi_1}{\partial x} = - E_1 \psi_2 + \xi \psi_1, \quad
      \frac{\partial \psi_2}{\partial x} = \sigma E_1^\ast \psi_1,
   } \\*[9pt] {\displaystyle
      \frac{\partial \psi_1}{\partial z} 
         = i \left( \sigma |E_1|^2 + \frac{\xi^2}{2} \right) \psi_1
           - i \left( d E_1 / d x + \xi E_1 \right) \psi_2, \quad
   } \\*[9pt] {\displaystyle
      \frac{\partial \psi_2}{\partial z} 
         = -i \left( \sigma |E_1|^2 + \frac{\xi^2}{2} \right) \psi_2
           - i \sigma \left( d E_1^\ast / d x 
                              - \xi E_1^\ast \right) \psi_1, \quad
   } \\*[9pt] {\displaystyle
      \frac{\partial \psi_3}{\partial x} = 0, \quad
      \frac{\partial \psi_3}{\partial z} 
         = - i \frac{\xi^2}{2} \psi_3.
   } \end{array}
\end{equation}
Here $\xi$ is the complex parameter which implicitly defines the soliton amplitude, width, and speed of its motion across the lattice. The functions $\psi_n$ can be found analytically for lattices with flat phase ($C=0$) profiles~\cite{Desyatnikov:2004:ProcNLGW+}. In the general case of lattices with arbitrary phase structure ($C \ne 0$), we can still derive a number of key analytical relations.
We note that the equations for $\psi_1$ and $\psi_2$ in~\reqt{psi} are uncoupled from $\psi_3$ and their general solution can be represented as a sum of two eigenmodes. With no loss of generality, we choose the eigenmodes which profiles remain stationary along the $z$ direction, similar to the underlying lattice, $\psi_{1,2}^{(j)}(x,z) = \psi_{1,2}^{(j)}(x) \exp(i \gamma_{1,2}^{(j)} z)$, where $j=1,2$ is the eigenmode index. The propagation constants are found as 
$\gamma_{1}^{(j)} 
= \gamma_{2}^{(j)} + \beta_1 
= \beta_1/2 
\pm \left[ (-\beta_1 + 2 \sigma |E_1|^2 + \xi^2)^2 
\right. + 4 \sigma ( d E_1 / d x + \xi E_1 ) \left. ( d E_1^\ast / d x - \xi E_1^\ast )
     \right]^{1/2}$, 
and $\gamma_3 = -\xi^2 / 2$. 
There is a specific relation between the amplitudes of two components,
$\psi_{1}^{(j)} / \psi_{2}^{(j)} 
= ( d E_1 / d x + \xi E_1 ) / ( -\gamma_1^{(j)} + \xi |E_1|^2 + \xi^2/2 )$.
Using these constraints, we can find the profiles of eigenmodes by integrating Eqs.~\reqt{psi} numerically. Then, the solution of original Eqs.~\reqt{NLS} describing a composite soliton on a lattice is found as
\begin{equation} \leqt{soliton}
   \begin{array}{l} {\displaystyle
      \widetilde{E}_1(x,z) 
         = E_1 
         - 2 {\rm Re}(\xi) \psi_1 \psi_2^\ast D^{-1} ,
   } \\*[9pt] {\displaystyle
      \widetilde{E}_2(x,z) 
         = - 2 {\rm Re}(\xi) \psi_1 \psi_3^\ast D^{-1} ,
   } \end{array}
\end{equation}
where $E_1$ is the nonlinear lattice profile,
and $D = \sigma |\psi_1|^2 + |\psi_2|^2 + |\psi_3|^2$. 
We can reveal the physical origin of solutions associated with different eigenmodes $\psi^{(j)}$ by explicitly taking into account soliton evolution along the propagation direction $z$, and write its profile in the following equivalent form,
\begin{equation} \leqt{sech}
   \begin{array}{l} {\displaystyle
       \widetilde{E}_2(x,z) 
          = - e^{i \beta_2^{(j)} z}{\rm Re}(\xi) 
            \left. \frac{\psi_1^{(j)} \psi_3^\ast}{ 
                           \sqrt{\sigma |\psi_1^{(j)}|^2 + |\psi_2^{(j)}|^2} |\psi_3|}
           \right|_{z=0}
   } \\*[9pt] {\displaystyle \qquad
        {\rm sech}\left[ {\rm Im}(\gamma_3 - \gamma_1^{(j)}) z + \chi(x) \right],
   } \end{array}
\end{equation}
where $\chi(x) = \left. {\rm ln}\left( \sqrt{\sigma |\psi_1^{(j)}|^2 + |\psi_2^{(j)}|^2} / |\psi_3| \right)\right|_{z=0}$, and the {\em soliton propagation constant} is $\beta_2^{(j)} = \gamma_1^{(j)} - \gamma_3^\ast$. It follows from Eq.~\reqt{sech} that the average {\em velocity of soliton motion across the lattice} is $v = {\rm lim}_{(x_2-x_1)\rightarrow \infty} {\rm Im}(\gamma_3 - \gamma_1) (x_2-x_1) / [ \chi(x_2) - \chi(x_1)]$. The initial soliton position $x_0$ at $z=0$ (or a fixed location of the stationary soliton) can be found from the condition $\chi(x_0)=0$, and it depends on the value of $\psi_3$, which is an arbitrary constant according to the last two equations in~\reqt{psi}.
Since the soliton is localized, $\beta_2$ should belong to a band-gap, where linear waves cannot propagate and become trapped at the self-induced nonlinear defect. We find that for stationary solitons ($v=0$), the propagation constants $\beta_2^{(j)}$ associated with two eigenmodes ($j=1,2$) of Eqs.~\reqt{psi} appear inside the total internal reflection or the Bragg-reflection gap, respectively, and the specific value of $\beta_2$ depends on the soliton parameter $\xi$. Accordingly, solitons can exist in both of these gaps in case of self-focusing ($\sigma=+1$) nonlinearity. On the other hand, only Bragg-gap solitons can form in media with self-defocusing ($\sigma=-1$) nonlinearity, as the denominator $D$ in Eq.~\reqt{soliton} becomes singular for modes in the other gap. 
These conclusions are in agreement with earlier numerical results for flat-phase nonlinear lattices~\cite{Desyatnikov:2003-153902:PRL}. As a matter of fact, these existence properties may seem to be the same as for solitons in fixed lattices (see Ref.~\cite{Pelinovsky:2004-36618:PRE}, and references therein), however there is a fundamental difference. For every value of the propagation constant, the position of stationary solitons with respect to the nonlinear lattice can be arbitrary depending on the choice of $\psi_3$ [cf. two bottom rows in Figs.~\rpict{focus}(a,c) and~\rpict{defoc}(a)], whereas only two specific locations on a period (one stable and one unstable) are possible in fixed lattices~\cite{Pelinovsky:2004-36618:PRE}. This indicates the absence of the self-induced Peierls-Nabarro potential, which can inhibit soliton motion through fixed lattices~\cite{Morandotti:1999-2726:PRL}.

\pict{fig02}{focus}{ 
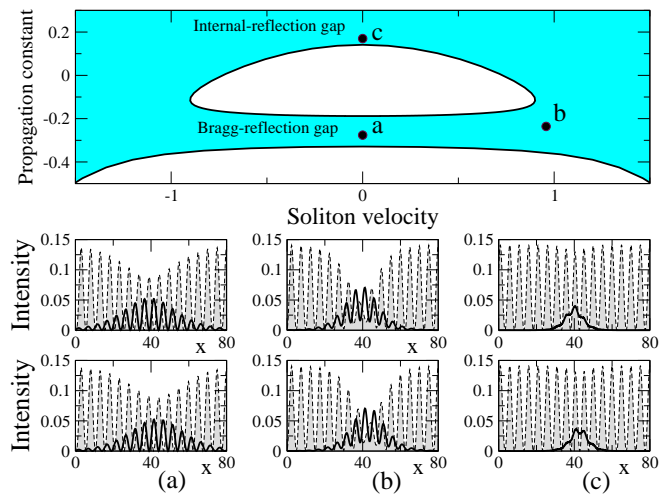
Top: Existence regions of the existence of stationary and moving composite band-gap solitons (shaded) in a self-focusing medium ($\sigma=+1$); the normalized propagation constant is $\beta_2 - 2 V_0$.
Bottom rows: (a-c)~Intensity profiles of the bright soliton (solid) and lattice (dashed and shaded) components corresponding to marked points in the top plot. Two rows illustrate different transverse positions. 
Lattice parameters correspond to Fig.~\rpict{lattice}(a).}

Even in the absence of the Peierls-Nabarro potential, certain conditions have to be  satisfied to sustain soliton motion through a nonlinear lattice. We note that the lattice is deformed at the soliton location, and if the corresponding defect is not created at the input the soliton mobility may be restricted. Even for optimal input conditions, we find that there appear fundamental limitations on the structure and mobility of solitons related to the properties of the band-gap spectrum.
We scan the full complex plane of soliton parameters $\xi$ and calculate how the existence regions of soliton propagation constants ($\beta_2$) change with the variation of their velocity. We first consider the properties of solitons in flat-phase lattices, and then use these results to identify phenomena due to lattice phase modulation.
In the self-focusing case ($\sigma=+1$), the existence regions expand at large velocities [see Fig.~\rpict{focus}(top)]. We obtain that the moving solitons can exist everywhere inside the {\em dynamical bandgaps} of the lattice spectrum, which edges are found from the linear Bloch-wave dispersion relations as $(- d\, \partial \beta_b / \partial K_b,\, \beta_b)$. 
The dynamical gap broadening is a fundamental phenomenon, however earlier studies of this effect were based on a simple coupled-mode theory which accounts for a single isolated gap (see Ref.~\cite{Conti:2001-36617:PRE}, and references therein). Our results describe a physical system with a non-trivial multi-gap spectrum, and we uncover the key effect of {\em gap merging} at large soliton velocities. In this regime, the propagation constant is detuned from the resonance with the lattice, and the large-amplitude soliton begins to locally erase the lattice, see Fig.~\rpict{focus}(b). 

\pict{fig03}{defoc}{ 
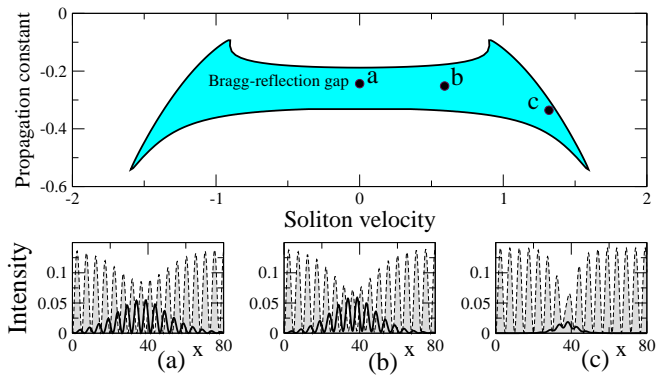
Soliton existence region and profiles in a self-defocusing medium ($\sigma=-1$). Notations are the same as in Fig.~\rpict{focus}, and lattice parameters correspond to Fig.~\rpict{lattice}(b).}

The features of moving bright solitons in media with self-defocusing nonlinearity are fundamentally different from the self-focusing case. 
At relatively small velocities, we observe the expansion of existence region which fully occupies the dynamic Bragg-reflection gap [see Fig.~\rpict{defoc}], however at larger velocities the existence region shrinks and completely disappears above a critical velocity. This happens because the soliton localization in self-defocusing media is only possible in the Bragg-reflection gap, which disappears through gap merging, as the effectiveness of scattering is reduced at larger velocities. Accordingly, only small-amplitude solitons are supported close to the existence boundary [see example in Fig.~\rpict{defoc}(c)]. 

\pict{fig04}{defocPhase}{ 
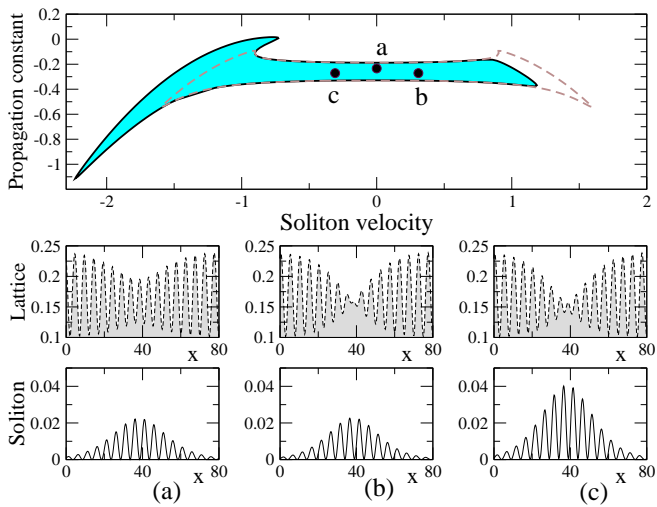
Top: Soliton existence region for a lattice with nontrivial phase modulation [Fig.~\rpict{lattice}(c)] in a self-defocusing medium shown with shading. Dashed line: boundary of the existence region of a trivial-phase lattice [Fig.~\rpict{lattice}(b)].
Bottom rows: Intensity profiles of the lattice and soliton components corresponding to the marked points in the top plot.
}

We now analyze the effect of lattice imbalance and associated phase modulation on the soliton dynamics. We have established that imbalanced lattices still have symmetric intensity profiles [see Eq.~\reqt{lattice} and Fig.~\rpict{lattice}(c)]. Since the coupling of spatially localized and periodic components is phase-insensitive, the left and right propagation directions are exactly equivalent for linear (small-amplitude) wave packets in the second component. However, at larger wave amplitudes, i.e. in the regime of soliton formation, the lattice becomes locally deformed and this process does depend on its phase, resulting in strong differences between the left- and right- moving solitons [cf. Figs.~\rpict{defocPhase}(b) and~(c)]. This is clearly indicated by a dramatic change in the soliton existence region in a self-defocusing medium [Fig.~\rpict{defocPhase}(top)], which becomes strongly asymmetric. For comparison, in the same plot we show with the dashed line the boundary of the existence region for a flat-phase lattice. We see that the existence regions exactly coincide and cover the whole dynamic Bragg-reflection gap up to a critical velocity. At large velocities, the gap-merging effect comes into play and existence region critically depends on the lattice phase: solitons can travel at much large velocities in the direction defined by the lattice phase gradient. 
We stress that this is a nontrivial result, which
has no analogs for fixed periodic structures.

In conclusion, we have described the dynamics of solitons in deformable lattices which are induced by periodic waves in a Kerr-type nonlinear medium. We have used exact analytical solutions to demonstrate that, even in the absence of the Peierls-Nabarro potential, the soliton mobility can be restricted due to other physical mechanisms, including the effect of band-gap merging. We have also suggested a new approach for controlling soliton motion by introducing a lattice excitation imbalance, while at the same time preserving the linear wave spectrum.

A.S. thanks Yuri Kivshar, Anton Desyatnikov, and Evgueny Doktorov for useful discussions and comments.

\end{sloppy}
\end{document}